\documentclass[natbib]{emulateapj}

\begin{document}

\title{Riding the Spiral Waves: Implications of Stellar Migration for the Properties of Galactic Disks}

\author{Rok Ro\v{s}kar\altaffilmark{1}, Victor
P. Debattista\altaffilmark{2}, Thomas R. Quinn\altaffilmark{1}, 
Gregory S. Stinson\altaffilmark{3}, James Wadsley\altaffilmark{3}}

\altaffiltext{1}{Astronomy Department, University of Washington, Box
351580, Seattle, WA 98195, USA {\tt
roskar;trq@astro.washington.edu}}
\altaffiltext{2}{RCUK Fellow at Centre for Astrophysics, University of Central
Lancashire, Preston, PR1 2HE, UK {\tt vpdebattista@uclan.ac.uk}}
\altaffiltext{3}{Department of Physics and Astronomy, McMaster University, 
Hamilton, ON, L8S 4M1, Canada {\tt stinson;wadsley@mcmaster.ca}}

\begin{abstract}

Stars in disks of spiral galaxies are usually assumed to remain roughly at their birth radii. 
This assumption is built into decades of modelling of the 
evolution of stellar populations in our own Galaxy and in external systems.  
We present results from self-consistent high-resolution $N$-body + Smooth 
Particle Hydrodynamics simulations of disk formation, in which stars migrate 
across significant galactocentric distances due to resonant scattering with transient spiral arms, 
while preserving their circular orbits. We investigate the implications 
of such migrations for observed stellar populations. Radial migration provides an 
explanation for the observed flatness and spread in the age-metallicity 
relation and the relative lack of metal poor stars in the solar neighborhood. 
The presence of radial migration also prompts rethinking of interpretations 
of extra-galactic stellar population data, especially for determinations of star formation histories. 

\end{abstract}


\section{Introduction}

In a universe where mass assembles by accretion of progressively larger constituents, 
thin disks of galaxies form during quiescent evolution following the major merging epoch
\citep{brook:2004,Robertson:2006}. The inner parts, with lower angular momentum 
assemble first, followed by the higher angular momentum components, resulting in 
``inside-out'' build up of disk material \citep{Larson:1976,white:1991,Munoz-Mateos:2007}. 
Because gas can efficiently dissipate energy, its motion around the galaxy is mostly circular. 
Stars born in such gas disks are initially on nearly circular orbits, but their infant kinematic state is  
highly fragile; non-axisymetric perturbations such as bars or spiral arms readily drive 
the orbits away from circular. Despite the increase in eccentricity, mean orbital radii are
assumed to remain relatively constant. As a result, interpretations of Galactic and 
extra-galactic stellar population observations invariably make the fundamental 
assumption that stars remain at roughly their birth radii throughout their lives. 
Consequently, this assumption is built into the vast majority of models of formation 
and evolution of galactic disks over the past few decades \citep[e.g.][]{Tinsley:1975,
Matteucci:1989, Carigi:1996,Chiappini:1997,Boissier:1999}.

Recent theoretical and observational evidence challenges this static picture. 
Spirals have long been known to be an important source of
kinematic heating in galactic disks, gradually increasing the eccentricities of 
stellar orbits \citep{Jenkins:1990}. However, if a star is 
caught in the corotation resonance of a transient spiral it may move inward
or outward in radius while preserving the circularity of its orbit 
\citep[][hereafter SB02]{Sellwood:2002}. 
Transient spirals with a wide range of pattern speeds are present throughout the 
evolution of the disk. Different pattern speeds result in spatially distinct locations of corotation resonances, allowing a star to undergo a series of resonant encounters in its 
lifetime, riding the spiral waves across large portions of the galaxy while retaining a 
mostly circular orbit. Stars born in-situ may therefore remain kinematically 
indistinguishable from those that have come across the galaxy. 

In \citet{Roskar:2008} we presented first results from our $N$-body Smooth
Particle Hydrodynamics (SPH) simulations of disk galaxy formation in which 
stars migrated radially due to their resonant interactions with transient spirals. 
These migrations yielded radial stellar population gradient predictions, 
which have recently been observed in resolved-star \citep{de-Jong:2007}, and
integrated-light studies \citep{Azzollini:2008, Bakos:2008}. 
In conjunction with our models, these observational data 
strongly imply that radial migration is an important process in 
observed galactic disks. In this Letter, we present further analysis of the simulation 
presented in \citet{Roskar:2008}, focusing on the wide-ranging implications of radial 
migrations for the observable properties of stellar populations.  

\section{Simulation}

The initial conditions are created as in \citet{Kaufmann:2006th}, and are designed to 
mimic the quiescent stage following a last major merger when thin disk 
formation commences. The system consists of a spherical dark 
matter NFW halo \citep{Navarro:1997aa} in which we embed a spherical 
halo of gas with the same initial profile. The gas halo is 
initially in hydrostatic equilibrium. The total mass of the system is $10^{12}$M$_{\odot}$, 
analogous to a Milky Way-type spiral galaxy, with the baryons contributing
10\% of the mass. We resolve the system with $10^6$ particles per component, 
resulting in initial mass resolutions of $10^6$M$_{\odot}$ and $10^5$M$_{\odot}$ 
for dark matter and gas respectively. Star particles form with masses that are a fraction 
of the gas particle mass, resulting in typical star masses  
around 3$\times$10$^4$M$_{\odot}$. To form a rotationally-supported 
disk, we also impart angular momentum to the gas component 
corresponding to a spin parameter value $\lambda$=0.039, as motivated by 
cosmological $N$-body experiments \citep[e.g.][]{Bullock:2001}. We evolve the system 
using the $N$-body + Smooth Particle Hydrodynamics code 
\textsc{gasoline} \citep{Wadsley:2004mb} for 10 Gyr.

Once the simulation begins, the gas cools and
collapses to the center of the halo, forming a thin rotating disk from the inside-out. 
When the gas reaches densities and temperatures which allow for star formation, 
the star formation and supernova (SN) feedback cycles are initiated
\citep{stinson:2006aa}. Since our disks form 
without any \textit{a-priori} assumptions about the 
interstellar medium (ISM) or the stellar populations present
in the disk, we can self-consistently follow the evolution of their properties as the disk grows.  
Transient spirals cause radial redistribution of 
stellar material in a manner analogous to the mechanism presented in SB02. 
Due to space limitations, a detailed discussion of these processes is deferred until 
a future paper. 

Our simulations do not account for the effects of evolution in a full cosmological context, but
in the standard $\Lambda$CDM paradigm mergers are significantly more important at early
epochs. The simulation presented here is therefore designed to mimic the 
quiescent phase of disk galaxy formation during which the thin disk forms. 
The increased resolution gained by stepping outside of the cosmological context allows us 
to isolate the effects of key dynamical processes, which determine multiple 
observational characteristics. Despite these simplifications, physical properties of our 
model such as the rotation curve, star formation rate, disk scale-length, 
and disk mass fraction, are analogous to those of observed systems. Pre-enrichment 
of merger fragments would not change the details of 
disk build-up, but simply offset the metallicity distribution. 

\section{Radial Migration and its Implications}

Models of galactic chemical evolution have been enormously successful in explaining
the properties of stars in our solar neighborhood \citep{Matteucci:1989,Carigi:1996,
Chiappini:1997,Boissier:1999}. One example of such model 
results is the age-metallicity relationship (AMR). The AMR is expected to arise
due to progressive enrichment of the ISM through stellar feedback. 
Stars of the same age in the same 
general region of the galaxy are therefore expected to have similar metallicities. 
Indeed, early determinations of the AMR confirmed that
the mean trend of stars in the solar neighborhood is toward lower metallicity with increasing age 
\citep{Twarog:1980}. Models, which assume that stars remain where 
they are born and return their nucleosynthetic yields to their local ISM, 
typically successfully reproduce this trend. However, evidence suggests that 
a large amount of scatter is present in the AMR of 
field stars \citep{Edvardsson:1993,Nordstrom:2004} and open clusters \citep{Friel:2002}, 
meaning that stars of the same age 
are observed to have a wide range of metallicities. 
In the framework of a galactic disk where 
radial annuli are self-enriched, homogeneous, and isolated from one another, old
metal-rich stars or very young metal-poor stars are an impossibility.  

Allowing stars from slightly different regions of the disk to enter the 
local sample due to the eccentricity of stellar 
orbits has been considered as a possible explanation, but can 
account only for up to 50\% of the observed scatter \citep{Nordstrom:2004,Binney:2007}. 
The large amount of scatter in the AMR therefore poses 
an important challenge for models of disk formation and should be considered 
as large a constraint as the mean trend \citep{Carraro:1998}.
The unexpected degree of scatter implies that stars in the 
solar neighborhood were either formed from a highly inhomogeneous ISM or have come
from wildly different parts of the galaxy. 

In our simulation, the latter option offers an enticing solution. In the left-most panel of 
Figure~\ref{fig:solar_neighborhood} we show the distribution of birth
radii for stars, which at the end of the simulation are on nearly circular orbits within an 
annulus analogous to the solar radius. We define
`solar radius' as a general region of the disk interior to the disk break, but 
approximately 2-3 scalelengths from the center. In our simulated galaxy this region is 
between 7-9 kpc (indicated by dashed lines in
Figure~\ref{fig:solar_neighborhood}). The black line represents
all stars while the blue and red lines show the distributions of metal poor 
([Fe/H] $< -0.3$) and metal rich ([Fe/H] $> -0.1$) stars respectively. 
Roughly 50\% of all ``solar neighborhood'' stars 
have come from elsewhere, primarily from the disk interior. 
Interestingly, some metal poor stars have been scattered into the solar
neighborhood from the outer part of the disk. Such migration has recently been inferred from  
observational data \citep{Haywood:2008}. Metal-rich stars, like our Sun, could have
originated almost anywhere in the Galaxy.

\subsection{Age-Metallicity Relationship}

%
%

\begin{figure*}
\centering
\plotone{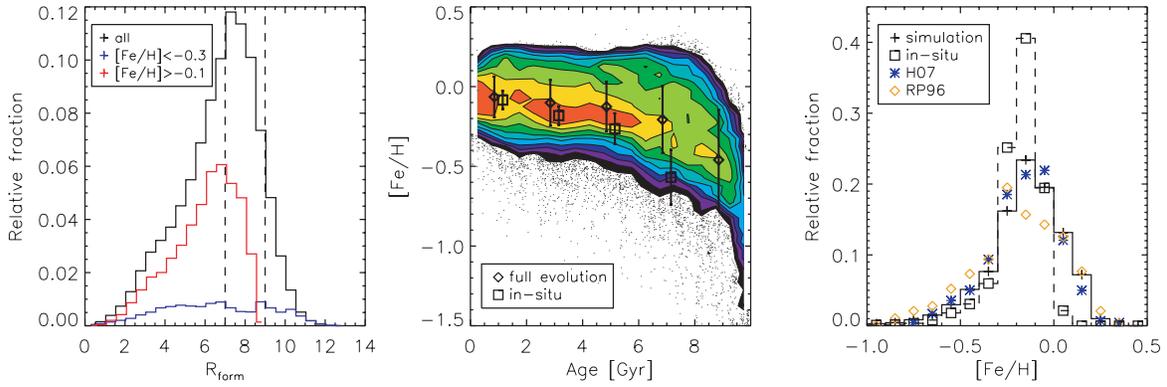}
\caption{Properties of stars in the solar neighborhood. 
{\bf Left:} Histogram of birth radii for stars that end up in the solar neighborhood 
on nearly circular orbits. The black, red, and blue lines represent all, metal-rich, 
and metal-poor stars respectively. 
{\bf Middle:} The age-metallicity relation: color 
contours represent relative particle densities where point density is high. Diamonds and 
error bars indicate mean values and dispersion respectively. Squares show
the AMR if stars are assumed to remain in-situ. A small horizontal 
offset is applied to the two sets of symbols for clarity.
{\bf Right:} Metallicity distribution function (MDF): the simulated distribution
is shown with the solid black histogram; diamonds and asterisks show data from 
\cite{Rocha-pinto:1996} and \cite{Holmberg:2007} respectively. The dashed 
histogram is the MDF if stars are assumed to remain in-situ. 
}
\label{fig:solar_neighborhood}
\end{figure*}

It has previously been suggested that such large radial migrations 
could account for the scatter in the observed 
AMR (SB02, \citealt{Binney:2007}), but our models allow us to 
examine the implications of migrations in a self-consistent simulation
of a growing and star-forming galactic disk. 
We show the AMR as determined from our simulation in the middle 
panel of Figure~\ref{fig:solar_neighborhood}. Colors indicate relative particle densities for 
clarity in high-density regions. Diamonds and error bars indicate the means and standard
deviations in several broad age bins.  With the exception of the oldest age bin, the AMR 
is fairly flat. The relatively high number of very old ($> 9$ Gyr) extremely metal-poor 
stars is partly a remnant of our initial configuration, which includes only 
pristine, metal-free gas. This is likely to be different in a cosmological setting where
a fraction of the gas is accreted from subhaloes that have already formed stars
and been enriched by stellar feedback. Also, note that we assume each star is born with the
mean metallicity of the ISM at its birth radius so that we can more easily isolate the effects of
radial migration on the various observational relations.

To illustrate the impact of the dynamical 
evolution of the disk, we plot the AMR in these same age bins by using only stars that 
actually formed in the solar neighborhood. This latter relation is analogous to that derived
by models which assume a dynamically non-evolving disk, essentially only depending on 
the nucleosynthetic yields and star formation and gas infall rates at the solar circle. 
In comparison to the full simulation, 
the in-situ relation shows a much tighter 
correlation between metallicity and age. Inclusion of full dynamical modeling 
significantly flattens the AMR and increases its dispersion by at least a factor of two. 
We note that although our simulations are not
designed to reproduce Milky Way observables, they successfully reproduce the 
major qualitative features of the observations: the observed AMR is indeed relatively flat with a  
high degree of scatter \citep{Edvardsson:1993,Nordstrom:2004}, 
which gradually increases with age \citep{Haywood:2008}. 
This agreement strongly suggests that a large amount of radial migration
is the missing piece of the AMR puzzle. 
 
\subsection{Metallicity Distribution Function} 
 
If radial migration is important for the AMR, then it should also leave a significant imprint
on the metallicity distribution function (MDF), another key observational constraint for
models of Galactic chemical evolution. The simple, closed-box 
chemical evolution picture where no material is allowed
to enter or leave a given radial annulus has long been recognized as inadequate
at explaining the local MDF 
\citep{van-den-Bergh:1962}. 
However, the dearth in the relative number of low-metallicity stars, 
known as the G-dwarf problem, is usually explained by allowing for the inflow of metal-poor gas 
\citep{Larson:1974,Lynden-Bell:1975}. Gas infall ``solves'' the G-dwarf problem
because it allows for prolonged star formation, which results in a relatively high number
of metal-rich stars.

Our simulated MDF is shown with diamonds in  
the right panel of Figure~\ref{fig:solar_neighborhood} compared to observational data
from \citet{Rocha-pinto:1996} and \citet{Holmberg:2007}, represented by 
diamonds and asterisks respectively.\footnote{Note that to shift the model quantities 
into the metallicity range relevant for the Milky Way, we apply an offset of +0.2 
dex to the simulated metallicities. We show the observed distributions not to 
claim quantitative agreement, but simply to illustrate the ramifications of 
dynamical effects.}
We also plot the MDF of stars formed only in the solar neighborhood (shown with squares).
Compared to the observed MDF, the in-situ distribution is much narrower. 
Radial migration introduces metal-poor and metal-rich stars into the 
solar neighborhood from other parts of the disk, thereby broadening the MDF. 

Unlike the solar neighborhood AMR, which is highly dispersed and therefore difficult
to use as a robust guide for Galactic chemical evolution models, the MDF represents 
a much more stringent constraint. It is clear from Figure~\ref{fig:solar_neighborhood}
that radial migration may significantly alter the nature of this fundamental property
of the solar neighborhood.

\subsection{Metallicity Gradients}

%
%

\begin{figure}
\centering
\plotone{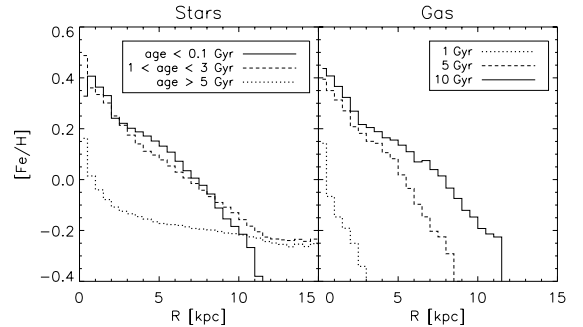}
\caption{
{\bf Left:}
Azimuthally averaged metallicity profile at 10 Gyr for stars in different age bins (young, 
intermediate age, and old stars shown with solid, dashed, and dotted lines respectively).  
{\bf Right:}
Azimuthally averaged metallicity profile of the gas at different times in the simulation. 
}
\label{fig:metallicity}
\end{figure}

A third major constraint on models of disk formation and evolution is the radial 
metallicity gradient.
If stars are assumed to remain in-situ, then stellar tracers of 
different ages should provide information about the ISM metallicity from 
which they spawned. 
In the left-hand panel of Figure~\ref{fig:metallicity} we show the metallicity gradient as 
determined from stellar tracers of different ages. The slope of the gradient
decreases with increasing age of stellar population. In most observational
studies, such a trend would be interpreted to mean that the slope of the ISM metallicity was shallower 
in the past \citep{Maciel:2005}. In the right-hand panel of Figure~\ref{fig:metallicity} 
we show the actual metallicity gradient in the ISM at different times in the 
simulation. Rather than being flatter in the past, as implied by the 
flat gradient of old stars, the ISM gradient was actually steeper and
evolved in the opposite sense to that indicated by the stellar tracers. Radial migration
causes more mixing in older populations, creating the appearance of flatter gradients 
at early times and leading to a decoupling of stellar properties from their birth ISM. 

\subsection{Star Formation History}

%
%

\begin{figure}
\centering
\plotone{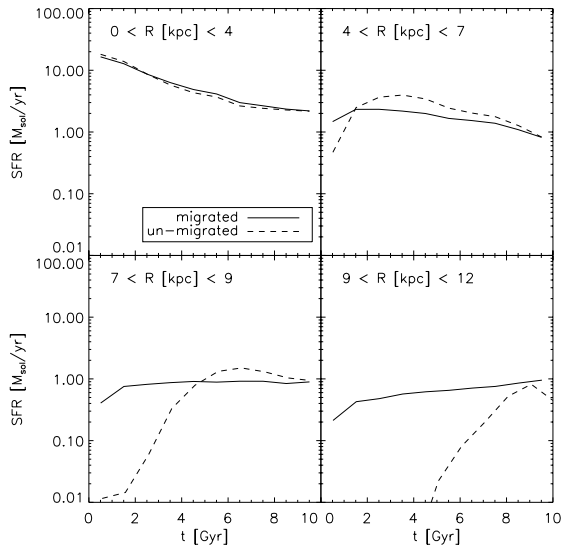}
\caption{
Star formation history in several broad radial bins. 
}
\label{fig:sfh}
\end{figure}

Implications of radial migration can be extended beyond our own Galaxy to 
extragalactic studies. In recent years, state-of-the-art 
observations of resolved stellar populations  have begun to probe 
detailed, spatially-resolved star formation histories in disks of external
galaxies \citep[e.g.][]{Barker:2007a, de-Jong:2007, Williams:2008, Gogarten:2008}. 
Typically, observed properties of stars in a given region of a 
disk are used to construct a color-magnitude 
diagram, which in turn is used to constrain stellar population 
synthesis models to yield a best-fit star formation history and therefore 
provide information about the formation of the disk \citep{Dolphin:2002}. 
Again, the underlying assumption is that stars observed today are found
close to the radius of their respective birthplaces.

In Figure~\ref{fig:sfh} we show the `migrated' (solid) and `un-migrated' (dashed) star 
formation history (SFH) in our simulation for several broad radial bins.   
The `migrated' SFH is constructed by considering the distribution of ages 
in the desired region of the disk at the final snapshot of the simulation, much like an observer would 
do given a snapshot of a present day galactic disk. The `un-migrated' SFH is 
the actual SFH, recovered by 
considering the times of formation for all stars formed within a given radial bin, regardless
of the final radius to which they eventually migrated. We show here all stars regardless 
of the eccentricity of their orbits, but our conclusions remain the same even if we restrict our
analysis only to stars on nearly circular orbits.

The discrepancy between the `migrated' and
`un-migrated' SFH is large, especially for the outermost regions of the disk. For example, the `solar
radius' bin (7 - 9 kpc) could erroneously be deduced to have a flat SFR for the past $\sim 8$ Gyr, 
when it actually didn't reach current SFR levels until just $\sim 5$ Gyr ago. 
Migration must therefore be accounted for even in extragalactic studies of stellar populations. 

\section{Conclusions}

We have shown that radial migration caused by resonant interactions of stars with transient
spirals has significant repercussions on the observed properties of a wide range of stellar
population systems. The range of implications and possible solutions to outstanding problems
given by radial migration is appealing. Further work is required to characterize fully the 
degrees and timescales of migration in spiral galaxies of different physical characteristics. On
the other hand, future observational efforts such as RAVE, GAIA and LSST will
provide intricate data sets that should further constrain the radial migration process in the 
Milky Way. Ongoing and future HST projects such as 
the ANGST and GHOSTS surveys, will require radial migration models in order to
reconstruct the SFH from resolved-star observations of nearby spiral disks and thereby 
learn about the formation history of our galactic neighborhood. 

\acknowledgments We would like to thank T. Kaufmann for the use of his initial 
conditions code. R.R. thanks C. Brook, J. Dalcanton and B. Gibson for useful discussions.
This research was supported in part by the NSF
through TeraGrid resources provided by TACC 
and PSC.  R. R. and T. R. Q. were supported by 
the NSF ITR grant PHY-0205413 at the University of Washington.
V. P. D. was supported by an RCUK Fellowship at the University of Central Lancashire.
R. R. acknowledges support for a visit to the
University of Central Lancashire from a Livesey Award Grant held by V. P. D.

\bibliographystyle{apj}


\begin{thebibliography}{40}
\expandafter\ifx\csname natexlab\endcsname\relax\def\natexlab#1{#1}\fi

\bibitem[{{Azzollini} {et~al.}(2008){Azzollini}, {Trujillo}, \&
  {Beckman}}]{Azzollini:2008}
{Azzollini}, R., {Trujillo}, I., \& {Beckman}, J.~E. 2008, \apjl, 679, L69

\bibitem[{{Bakos} {et~al.}(2008){Bakos}, {Trujillo}, \& {Pohlen}}]{Bakos:2008}
{Bakos}, J., {Trujillo}, I., \& {Pohlen}, M. 2008, ArXiv e-prints, 807

\bibitem[{{Barker} {et~al.}(2007){Barker}, {Sarajedini}, {Geisler}, {Harding},
  \& {Schommer}}]{Barker:2007a}
{Barker}, M.~K., {Sarajedini}, A., {Geisler}, D., {Harding}, P., \& {Schommer},
  R. 2007, \aj, 133, 1138

\bibitem[{{Binney}(2007)}]{Binney:2007}
{Binney}, J. 2007, {Dynamics of Disks} (Island Universes - Structure and
  Evolution of Disk Galaxies), 67--+

\bibitem[{{Boissier} \& {Prantzos}(1999)}]{Boissier:1999}
{Boissier}, S. \& {Prantzos}, N. 1999, \mnras, 307, 857

\bibitem[{{Brook} {et~al.}(2004){Brook}, {Kawata}, {Gibson}, \&
  {Freeman}}]{brook:2004}
{Brook}, C.~B., {Kawata}, D., {Gibson}, B.~K., \& {Freeman}, K.~C. 2004, \apj,
  612, 894

\bibitem[{{Bullock} {et~al.}(2001){Bullock}, {Dekel}, {Kolatt}, {Kravtsov},
  {Klypin}, {Porciani}, \& {Primack}}]{Bullock:2001}
{Bullock}, J.~S., {Dekel}, A., {Kolatt}, T.~S., {Kravtsov}, A.~V., {Klypin},
  A.~A., {Porciani}, C., \& {Primack}, J.~R. 2001, \apj, 555, 240

\bibitem[{{Carigi}(1996)}]{Carigi:1996}
{Carigi}, L. 1996, RevMexAA, 32, 179

\bibitem[{{Carraro} {et~al.}(1998){Carraro}, {Ng}, \&
  {Portinari}}]{Carraro:1998}
{Carraro}, G., {Ng}, Y.~K., \& {Portinari}, L. 1998, \mnras, 296, 1045

\bibitem[{{Chiappini} {et~al.}(1997){Chiappini}, {Matteucci}, \&
  {Gratton}}]{Chiappini:1997}
{Chiappini}, C., {Matteucci}, F., \& {Gratton}, R. 1997, \apj, 477, 765

\bibitem[{{de Jong} {et~al.}(2007){de Jong}, {Seth}, {Radburn-Smith}, {Bell},
  {Brown}, {Bullock}, {Courteau}, {Dalcanton}, {Ferguson}, {Goudfrooij},
  {Holfeltz}, {Holwerda}, {Purcell}, {Sick}, \& {Zucker}}]{de-Jong:2007}
{de Jong et~al.} 2007, \apjl, 667, L49

\bibitem[{{Dolphin}(2002)}]{Dolphin:2002}
{Dolphin}, A.~E. 2002, \mnras, 332, 91

\bibitem[{{Edvardsson} {et~al.}(1993){Edvardsson}, {Andersen}, {Gustafsson},
  {Lambert}, {Nissen}, \& {Tomkin}}]{Edvardsson:1993}
{Edvardsson}, B., {Andersen}, J., {Gustafsson}, B., {Lambert}, D.~L., {Nissen},
  P.~E., \& {Tomkin}, J. 1993, \aap, 275, 101

\bibitem[{{Friel} {et~al.}(2002){Friel}, {Janes}, {Tavarez}, {Scott},
  {Katsanis}, {Lotz}, {Hong}, \& {Miller}}]{Friel:2002}
{Friel et al.} 2002, \aj, 124, 2693

\bibitem[{{Gogarten} {et~al.}(2008)}]{Gogarten:2008}
{Gogarten et~al.} 2008, \apj, submitted

\bibitem[{{Haywood}(2008)}]{Haywood:2008}
{Haywood}, M. 2008, ArXiv e-prints, 805

\bibitem[{{Holmberg} {et~al.}(2007){Holmberg}, {Nordstr{\"o}m}, \&
  {Andersen}}]{Holmberg:2007}
{Holmberg}, J., {Nordstr{\"o}m}, B., \& {Andersen}, J. 2007, \aap, 475, 519

\bibitem[{{Jenkins} \& {Binney}(1990)}]{Jenkins:1990}
{Jenkins}, A. \& {Binney}, J. 1990, \mnras, 245, 305

\bibitem[{{Kaufmann} {et~al.}(2006){Kaufmann}, {Mayer}, {Wadsley}, {Stadel}, \&
  {Moore}}]{Kaufmann:2006th}
{Kaufmann}, T., {Mayer}, L., {Wadsley}, J., {Stadel}, J., \& {Moore}, B. 2006,
  \mnras, 370, 1612

\bibitem[{{Larson}(1974)}]{Larson:1974}
{Larson}, R.~B. 1974, \mnras, 166, 585

\bibitem[{{Larson}(1976)}]{Larson:1976}
---. 1976, \mnras, 176, 31

\bibitem[{{Lynden-Bell}(1975)}]{Lynden-Bell:1975}
{Lynden-Bell}, D. 1975, Vistas in Astronomy, 19, 299

\bibitem[{{Maciel} {et~al.}(2005){Maciel}, {Lago}, \& {Costa}}]{Maciel:2005}
{Maciel}, W.~J., {Lago}, L.~G., \& {Costa}, R.~D.~D. 2005, \aap, 433, 127

\bibitem[{{Matteucci} \& {Francois}(1989)}]{Matteucci:1989}
{Matteucci}, F. \& {Francois}, P. 1989, \mnras, 239, 885

\bibitem[{{Mu{\~n}oz-Mateos} {et~al.}(2007){Mu{\~n}oz-Mateos}, {Gil de Paz},
  {Boissier}, {Zamorano}, {Jarrett}, {Gallego}, \&
  {Madore}}]{Munoz-Mateos:2007}
{Mu{\~n}oz-Mateos}, J.~C., {Gil de Paz}, A., {Boissier}, S., {Zamorano}, J.,
  {Jarrett}, T., {Gallego}, J., \& {Madore}, B.~F. 2007, \apj, 658, 1006

\bibitem[{{Navarro} {et~al.}(1997){Navarro}, {Frenk}, \&
  {White}}]{Navarro:1997aa}
{Navarro}, J.~F., {Frenk}, C.~S., \& {White}, S.~D.~M. 1997, \apj, 490, 493

\bibitem[{{Nordstr{\"o}m} {et~al.}(2004){Nordstr{\"o}m}, {Mayor}, {Andersen},
  {Holmberg}, {Pont}, {J{\o}rgensen}, {Olsen}, {Udry}, \&
  {Mowlavi}}]{Nordstrom:2004}
{Nordstr{\"o}m et al.} 2004, \aap, 418, 989

\bibitem[{{Robertson} {et~al.}(2006){Robertson}, {Bullock}, {Cox}, {Di Matteo},
  {Hernquist}, {Springel}, \& {Yoshida}}]{Robertson:2006}
{Robertson}, B., {Bullock}, J.~S., {Cox}, T.~J., {Di Matteo}, T., {Hernquist},
  L., {Springel}, V., \& {Yoshida}, N. 2006, \apj, 645, 986

\bibitem[{{Rocha-Pinto} \& {Maciel}(1996)}]{Rocha-pinto:1996}
{Rocha-Pinto}, H.~J. \& {Maciel}, W.~J. 1996, \mnras, 279, 447

\bibitem[{{Ro{\v s}kar} {et~al.}(2008){Ro{\v s}kar}, {Debattista}, {Stinson},
  {Quinn}, {Kaufmann}, \& {Wadsley}}]{Roskar:2008}
{Ro{\v s}kar}, R., {Debattista}, V.~P., {Stinson}, G.~S., {Quinn}, T.~R.,
  {Kaufmann}, T., \& {Wadsley}, J. 2008, \apjl, 675, L65

\bibitem[{{Sellwood} \& {Binney}(2002)}]{Sellwood:2002}
{Sellwood}, J.~A. \& {Binney}, J.~J. 2002, \mnras, 336, 785

\bibitem[{{Stinson} {et~al.}(2006){Stinson}, {Seth}, {Katz}, {Wadsley},
  {Governato}, \& {Quinn}}]{stinson:2006aa}
{Stinson}, G., {Seth}, A., {Katz}, N., {Wadsley}, J., {Governato}, F., \&
  {Quinn}, T. 2006, \mnras, 373, 1074

\bibitem[{{Tinsley}(1975)}]{Tinsley:1975}
{Tinsley}, B.~M. 1975, \apj, 197, 159

\bibitem[{{Twarog}(1980)}]{Twarog:1980}
{Twarog}, B.~A. 1980, \apj, 242, 242

\bibitem[{{van den Bergh}(1962)}]{van-den-Bergh:1962}
{van den Bergh}, S. 1962, \aj, 67, 486

\bibitem[{{Wadsley} {et~al.}(2004){Wadsley}, {Stadel}, \&
  {Quinn}}]{Wadsley:2004mb}
{Wadsley}, J.~W., {Stadel}, J., \& {Quinn}, T. 2004, New Astronomy, 9, 137

\bibitem[{{White} \& {Frenk}(1991)}]{white:1991}
{White}, S.~D.~M. \& {Frenk}, C.~S. 1991, \apj, 379, 52

\bibitem[{{Williams} {et~al.}(2008)}]{Williams:2008}
{Williams et~al.} 2008, \aj, submitted

\end{thebibliography}

\clearpage

\end{document}